\documentclass[onecolumn]{autart}

\usepackage[cmex10]{amsmath}
\usepackage{amscd,amsfonts,amsmath,amssymb,mathrsfs,pifont,stmaryrd,tipa,color}
\usepackage{array,cases,dsfont,graphicx,texdraw}
\usepackage{graphicx}
\usepackage{epsfig,fancyhdr} % for postscript graphics files

\newtheorem{Remark}{Remark}[section]

\newenvironment{Proof}{\noindent{\em Proof:\/}}{\hfill $\Box$\par}
\newtheorem{Theorem}{Theorem}[section]
\newtheorem{Lemma}{Lemma}[section]

\newtheorem{Assumption}{Assumption}

\newcommand{\col}{\hbox{col}}
\newcommand{\EQQ}{\begin{eqnarray*}}
\newcommand{\ENN}{\end{eqnarray*}}
\newcommand{\EQ}{\begin{eqnarray}}
\newcommand{\EN}{\end{eqnarray}}
\begin{document}

\begin{frontmatter}

\title{Output Based Adaptive Distributed Output Observer for Leader-follower Multiagent Systems\thanksref{footnoteinfo}} % Title, preferably not more

\thanks[footnoteinfo]{This work has been supported in part by the Research Grants
Council of the Hong Kong Special Administration Region under grant
No. 14201418, and in part by National Natural Science Foundation of
China under grant No. 61973260. Corresponding author: J. Huang. Tel.
+852-39438473. Fax +852-39436002.}

\author[Cai-Huang]{He Cai}\ead{caihe@scut.edu.cn},
\author[Cai-Huang]{Jie Huang}\ead{jhuang@mae.cuhk.edu.hk}.

\address[Cai-Huang]{Department of Mechanical and Automation
Engineering, The Chinese University of Hong Kong, Shatin, N.T., Hong
Kong.}


\begin{abstract}
The adaptive distributed observer approach has been an effective tool for synthesizing a distributed control law for solving
various control problems of leader-follower multiagent systems.
However, the existing adaptive distributed observer  needs to make use of the full state of the leader system.
This assumption not only precludes many practical applications in which only the output of the leader system is available, but also
leads to a high dimension observer. In this communique,
we propose an  adaptive distributed output observer  which only makes use of
the output of the leader system, and is thus more practical than the
state based adaptive distributed observer. Moreover, the dimension and the information
exchange among agents of the proposed adaptive distributed output observer can be significantly smaller than those of the state based adaptive distributed output observer.
\end{abstract}

\end{frontmatter}

\fancypagestyle{plain}

\section{Introduction}

The past decade has witnessed a significant advancement on the research of
multiagent systems \cite{chen14,jlm03,kc16,lzd14,om04,q09,rb08}.
A variety of approaches have been developed for handling various control problems. One of
the systematic and effective approaches for the control
of leader-follower multiagent systems is the so-called distributed observer approach \cite{ys12}, which
consists of two design steps. First,  a distributed observer is synthesized for the given leader system. This distributed observer will provide
the estimation of the leader's state
to each follower system satisfying the communication constraints. Second, based on the estimated
leader's state provided by the distributed observer,
a certainty equivalent control law is synthesized for each follower to achieve the control objective.
A typical example of the application of the distributed observer approach can be found in
\cite{fhrd18}, where the distributed observer was used to recover the reference signals for solving the distributed robust tracking problem of a leader-follower Euler-Lagrange multiagent system.

Nevertheless,
the distributed observer approach has one drawback in that it assumes that the control law of each follower knows the leader's system matrix, which may not be desirable in some practical applications.
To remove this assumption, the adaptive distributed observer approach
was further proposed  in \cite{clgh16}, which not only estimates the
state of the leader system but also the system matrix of the leader system.  As a result,
the assumption that all the followers know the leader's system matrix is removed, which enables the design of
a fully distributed control law. Some other variants on the adaptive distributed observer approach, such as
the  one that guarantees  finite time convergence and the one that deals with   multiple leaders, can be found in \cite{hdlr20,lzmz19}.

Both the existing distributed observers and adaptive distributed observers need to make use of the full state of the leader system.
This assumption not only precludes many practical applications in which only the output of the leader system is available, but also
leads to a high dimension observer. In this communique,
we propose an output based adaptive distributed output observer depending solely on the output of the leader system,
which is the only practical solution in case where the state of the leader is not available.
Moreover,
instead of estimating every entry of the leader's system and output matrices, we only estimate the
coefficients of the minimal polynomial of the leader's system matrix.
As a result,
both the dimension and the information
exchange among agents of the proposed output based adaptive distributed output observer
can be drastically smaller than those of the state based adaptive distributed output observer.

%The attitude control problem is  among the most important issues for
%spacecraft maneuver and robot manipulation \cite{wkd91,y88}.
%The leader-following attitude consensus problem for multiple rigid body systems
%is one of the
%fundamental problems for the cooperative control of multiagent systems, and has been studied by ourselves
% in \cite{ch14} using the distributed observer approach and in \cite{chifac17} using the adaptive distributed observer approach. It is also  extensively studied
%in \cite{lds14,r072,vh06}, to name just a few. As a direct
%application of our main result, we will synthesize a distributed
%control law to solve the leader-following attitude consensus problem
%by integrating
%the output based adaptive distributed observer with
%a purely decentralized control
%law invoking the
%certainty equivalence principle.
%It will be seen that the output
%based adaptive distributed  observer needs much less information of the leader system
%than the existing adaptive distributed observer
%and thus leads to a distributed control law with much lower dimension.

The rest of this communique is organized as follows. Notation and preliminaries
are summarized in Section \ref{snmp}. Section \ref{smr} presents the
design of the output based adaptive distributed output observer. A numerical example is given
in Section \ref{ss}. Section \ref{sc} concludes this communique.

\section{Notation and Mathematical Preliminaries}\label{snmp}

\subsection{Notation}
$\mathbb{R}$ and $\mathbb{C}$ denote the sets of real and complex numbers, respectively.
For $x\in \mathbb{C}$, $\Re(x)$ denotes the real part of $x$.
$1_N$ denotes an $N$ dimensional column vector whose components are all $1$.
$||x||$ denotes the Euclidean norm of a vector $x$.
$||A||$ denotes the Euclidean norm of a matrix $A$.
$A^T$ ($A^H$) denotes the (Hermitian) transpose of $A$.
For a square matrix $A$, $\sigma(A)$ denotes the spectrum of $A$,
$\Re(\sigma(A))$ denotes the set of the real parts of all the elements of $\sigma(A)$,
$\bar{\delta}_A=\max(\Re(\sigma(A)))$, $\underline{\delta}_A=\min(\Re(\sigma(A)))$,
and
$A>0$ $(A \geq 0)$ means $A$ is positive definite (positive semi-definite).
A matrix $A=[a_{ij}]\in \mathbb{R}^{n\times n}$ is called an $M$-matrix
if $a_{ij}\leq 0$ for $i\neq j$ and
$\underline{\delta}_A>0$.
For two hermitian matrices $X_1$ and $X_2$, $X_1\leq X_2$ means $X_2-X_1\geq 0$.
Given a time-varying matrix $A(t)$, if $||A(t)||\leq \beta e^{-\alpha t}$ for some
$\alpha,\beta>0$, then
$A(t)$ is said to decay to zero exponentially at the rate of $\alpha$.
$\otimes$ denotes the Kronecker product of matrices.
For $x_i\in \mathbb{R}^{n_i}$, $i=1,\dots,m$,
$\col(x_1,\dots,x_m)=[x_1^T,\dots,x_m^T]^T$.

%A digraph $\mathcal {G}=(\mathcal {V},\mathcal {E})$
%consists of a node set $\mathcal {V}=\{1,\dots,N\}$ and an
%edge set $\mathcal {E} = \{(i,j), i, j \in \mathcal {V}, i\neq j
%\}$.  An edge from node $i$ to node $j$ is denoted by $(i,j)$, and
%node $i$ is called the neighbor of node $j$. If the digraph
%$\mathcal{G}$ contains a sequence of edges in the form of $({i_1},
%{i_2}), ({i_2}, {i_3}), \dots, ({i_{k}}, {i_{k+1}})$, then the set
%$\{({i_1}, {i_2}), ({i_2}, {i_3}), \dots, ({i_{k}}, {i_{k+1}}) \}$
%is called a directed path of $\mathcal{G}$ from node ${i_1}$ to node ${i_{k+1}}$, and
%node ${i_{k+1}}$ is said to be reachable from node ${i_{1}}$.   A
%digraph is said to contain a spanning tree  if there exists a node $i$
%such that any other node is reachable from node $i$, and node $i$
%is called the root of the spanning tree.  A digraph
%$\mathcal{G}_s=(\mathcal{V}_s,\mathcal{E}_s)$ is called a subgraph
%of $\mathcal{G}=(\mathcal{V},\mathcal{E})$ if
%$\mathcal{V}_s\subseteq \mathcal {V}$ and $\mathcal{E}_s\subseteq
%\mathcal{E}\cap (\mathcal{V}_s\times \mathcal{V}_s)$.  The weighted adjacency
%matrix $\mathcal{A}=[a_{ij}]\in \mathbb{R}^{N\times N}$ of $\mathcal{G}$ is
%defined as $a_{ii}=0$; for $i\neq j$,
% $a_{ij} >0  \Leftrightarrow (j,i)\in \mathcal{E}$ and $a_{ij}=0$ otherwise.
% The Laplacian $\mathcal{L}=[l_{ij}]\in \mathbb{R}^{N\times N}$ of $\mathcal{G}$ is defined
% as $l_{ii} =\sum_{j=1}^Na_{ij} $; for $i\neq j$, $l_{ij} =-a_{ij}$.
%

 \subsection{Preliminaries}

First,  we  list two lemmas  for convenience of the readers.

\begin{Lemma}\label{cor2}
  Given a detectable pair $(C,A)$, where $A\in \mathbb{R}^{n\times n}$,
  $C\in \mathbb{R}^{m\times n}$, let $P>0$
  be the unique solution of the algebraic Riccati equation
  \begin{equation}
    PA^T+AP-PC^TCP+Q=0
  \end{equation}
 for some $Q>0$.
Let $\Phi=I_N\otimes A-\mu(F\otimes PC^TC)$
  where $F\in \mathbb{R}^{N\times N}$ satisfies
  $\underline{\delta}_F>0$. Then,
   $\Phi$ is Hurwitz if $\mu>\underline{\delta}_F^{-1}$.
\end{Lemma}

The proof of Lemma \ref{cor2} can be extracted from the proof of Theorem 2 of \cite{ys13}.

\begin{Lemma}\label{lemap1} (p.168, p.260, Theorem 11.2.1 of \cite{lr95})
  Consider the following equation
  \begin{equation}\label{lemap1.1}
    XDX-XA-A^HX-C=0
  \end{equation}
  where $A, C, D, X \in \mathbb{C}^{n\times n}$.
  A hermitian solution $X_{+}$ of \eqref{lemap1.1} is called maximal if $X\leq X_+$ for every
  hermitian solution $X$ of \eqref{lemap1.1}. $X_+$ as a function of $A$, $C$ and $D$ is expressed
  as $X_+=X_+(A,C,D)$. Let $\mathds{P}$ denote the set of all ordered triples $(A,C,D)$
  satisfying $D\geq 0$, $C=C^H$, $(A,D)$ is stabilizable, and \eqref{lemap1.1}
  admits hermitian solutions. Then, the maximal
  hermitian solution $X_{+}(A,C,D)$ of \eqref{lemap1.1} is a continuous function of $(A,C,D)\in \mathds{P}$.
\end{Lemma}

%\begin{Lemma}\label{lemap5}(Theorem 2.5.3 of \cite{horn})
%  A matrix $A$ is an $M$-matrix if and only if there exists $D=\hbox{diag}\{d_1,\dots,d_N\}>0$
%  such that $DA+A^TD>0$.
%\end{Lemma}

Next, we summarize the result from \cite{clgh16} regarding the adaptive distributed observer.
Consider the linear system described as follows:
\begin{subequations}\label{leader}
    \begin{align}
      \dot{v}_0 &= S_0v_0  \label{leader.1}\\
       y_0 &= C_0v_0 \label{leader.2}
      \end{align}
\end{subequations}
where $v_0\in \mathbb{R}^q$, $y_0\in \mathbb{R}^p$, $S_0\in \mathbb{R}^{q\times q}$ and $C_0\in \mathbb{R}^{p\times q}$ are constant matrices.

Let $\bar{\mathcal{G}}=(\bar{\mathcal{V}},\bar{\mathcal{E}})$ denote a
digraph\footnote{See Notation of \cite{clgh16} for a summary of graph notation.}  with
$\bar{\mathcal{V}}=\{0,1,\dots,N\}$.
Here the node $0$ is associated with the leader system \eqref{leader} and node $i$ is associated
with the $i$th follower. It is assumed that
the digraph $\bar{\mathcal{G}}=(\bar{\mathcal{V}},\bar{\mathcal{E}})$ satisfies the following assumption:

\begin{Assumption}\label{ass1}
The communication graph $\bar{\mathcal{G}}$ contains a spanning tree with the node $0$ as the root.
\end{Assumption}

\begin{Remark}
Assumption \ref{ass1} is a standard and necessary
assumption for the control of the leader-follower multiagent systems under static network topology.
Let $\bar{\mathcal{L}}$ be the Lapacian of $\bar{\mathcal{G}}$, and
$H$ consist of the last $N$ rows and the last $N$ columns of $\bar{\mathcal{L}}$.
Then,   by Lemma 4 of \cite{Hu1} or
Lemma 1 of \cite{ys12}, under Assumption \ref{ass1}, $-H$ is Hurwitz. Thus,
$\underline{\delta}_H>0$.
\end{Remark}

Given the system \eqref{leader} and the graph $\bar{\mathcal{G}}$,
we established the  adaptive distributed observer for the leader system in \cite{chwcica18} as follows:
\begin{subequations}\label{ado}
    \begin{align}
      \dot{S}_i&=\mu_s\sum_{j=0}^Na_{ij}(S_j-S_i) \label{ado.1}\\
      \dot{C}_i&=\mu_c\sum_{j=0}^Na_{ij}(C_j-C_i) \label{ado.2}\\
      \dot{v}_i&=S_iv_i+\mu_v\sum_{j=0}^Na_{ij}(v_j-v_i) \label{ado.3}
      \end{align}
\end{subequations}
where $S_i\in \mathbb{R}^{q\times q}$, $C_i\in \mathbb{R}^{p\times q}$, $v_i\in \mathbb{R}^{q}$, $\mu_s,\mu_c,\mu_v>0$.
Let $y_i=C_iv_i$, $\tilde{y}_i=y_i-y_0$, $\tilde{S}_i=S_i-S_0$, $\tilde{v}_i=v_i-v_0$, $\tilde{C}_i=C_i-C_0$, $i=1,\dots,N$.
Noting that $\tilde{y}_i=C_i\tilde{v}_i+\tilde{C}_iv_0$, by Lemma 2 of \cite{clgh16}, we can obtain the
following result:

\begin{Lemma} \label{lem1}
Given system \eqref{ado},  under Assumption \ref{ass1}, for all $v_i(0)\in \mathbb{R}^{q}$, $i=0,1,\dots,N$,
$S_i(0)\in \mathbb{R}^{q\times q}$,  $C_i(0)\in \mathbb{R}^{p\times q}$,
$i=1,\dots,N$,  we have
\begin{description}
  \item[(i)] for any $\mu_s,\mu_c>0$,
$S_i(t)$ and $C_i(t)$ exist for all $t\geq 0$ and satisfy
$\lim_{t\rightarrow\infty}\tilde{S}_i(t)=0$,
$\lim_{t\rightarrow\infty}\tilde{C}_i(t)=0$ exponentially at the rate of $\mu_s\underline{\delta}_H$,
$\mu_c\underline{\delta}_H$, respectively;
  \item[(ii)] if $\mu_s,\mu_c,\mu_v>\bar{\delta}_{S_0}\underline{\delta}_H^{-1}$,
then $y_i(t)$ exists for all $t\geq 0$ and satisfies
$\lim_{t\rightarrow\infty}\tilde{y}_i(t)=0$ exponentially.
\end{description}
\end{Lemma}

\begin{Remark}
The  adaptive distributed output observer (\ref{ado}) was first proposed in \cite{clgh16}  for the special case
where the matrix $C_0$ in \eqref{leader} is an identity matrix.
For this special case, $y_0 = v_0$. Thus, there is no need to  estimate $C_0$ in order  to recover $y_0$,
and the  adaptive distributed observer given in \cite{clgh16}  consists of only \eqref{ado.1} and \eqref{ado.3}.
Since  \eqref{ado} assumes that the state $v_0$ of the leader system is available,
it can be more precisely called state based adaptive distributed output observer.
\end{Remark}

%Under Assumption \ref{ass1}, for any
% $C_i(0)\in \mathbb{R}^{p\times q}$, $i=1,\dots,N$, we have
%\begin{description}
%  \item[(i)] for any $\mu_c>0$, $C_i(t)$ exists for all $t\geq 0$ and satisfies $\lim_{t\rightarrow\infty}\tilde{C}_i(t)=0$ exponentially at the rate of $\mu_c\underline{\delta}_H$;
%  \item[(ii)] if $\mu_c>\bar{\delta}_{S_0}\underline{\delta}_H^{-1}$, $\lim_{t\rightarrow\infty}\tilde{y}_i(t)=0$ exponentially.
%\end{description}

\section{Main Result}\label{smr}

In this section, we offer two significant improvements over \eqref{ado}.
First, \eqref{ado} needs the full state of the leader system.
But, in many practical applications, only the output of the leader system is available.
Thus, we will  propose  a so-called output based adaptive distributed output observer that
only relies on the output $y_0$ of the leader system.
Second, all the entries of $S_0$ and $C_0$
need to be estimated by each follower using \eqref{ado}.
In contrast, we
will show that it suffices to estimate the
coefficients of the minimal polynomial of $S_0$ instead of
all the entries of $S_0$ and $C_0$.  Thus, the proposed output based adaptive distributed output observer
can drastically reduce the dimension of the observer as well as the
information exchange among agents.

Suppose the minimal polynomial of $S_0$ is given by $s^{n}+\alpha_{0,1}s^{n-1}+\cdots+\alpha_{0,n-1}s+\alpha_{0,n}$.
Then, by the Cayley-Hamilton Theorem,
\begin{equation}
    S_0^n+\alpha_{0,1}S_0^{n-1}+\dots+\alpha_{0,n-1}S_0+\alpha_{0,n}I_q=0.
\end{equation}
By \eqref{leader}, for $k=0,1,\dots,n$,
\begin{equation}
    y_0^{(k)}=C_0S_0^{k}v_0.
\end{equation}
Then
\begin{equation}
\begin{aligned}
    &y_0^{(n)}+\alpha_{0,1}y_0^{(n-1)}+\dots+\alpha_{0,n-1}y_0^{(1)}+\alpha_{0,n}y_0\\
    =&C_0S_0^nv_0+\alpha_{0,1}C_0S_0^{n-1}v_0+\cdots\\
    &+\alpha_{0,n-1}C_0S_0v_0+\alpha_{0,n}C_0v_0\\
    =&C_0(S_0^n+\alpha_{0,1}S_0^{n-1}+\dots+\alpha_{0,n-1}S_0+\alpha_{0,n}I_q)v_0=0.
\end{aligned}
\end{equation}
Let $\zeta_{0}=\col(y_0,y_0^{(1)},\dots,y_0^{(n-1)})\in \mathbb{R}^{pn}$. Then,
\begin{equation}\label{exoeq}
\begin{aligned}
    \dot{\zeta}_{0}&=\left[\left(
                  \begin{array}{cccc}
                    0 &  &  &  \\
                    \vdots &  & I_{n-1} &  \\
                    0 &  &  & \\
                    -\alpha_{0,n} & \cdots & -\alpha_{0,2} & -\alpha_{0,1} \\
                  \end{array}
                \right)
    \otimes I_p\right]\zeta_{0}\\
    & \triangleq \mathcal{S}_0 \zeta_{0}\\
   y_0&=\left[\left(
            \begin{array}{cccc}
              1 & 0 & \cdots & 0 \\
            \end{array}
          \right)
    \otimes I_p\right]\zeta_0 \triangleq \mathcal{C}_0 \zeta_{0}.
\end{aligned}
\end{equation}
Since $(\mathcal{C}_0,\mathcal{S}_0)$ is observable, let
$\mathcal{P}_0>0$ be the unique solution of
the algebraic Riccati equation
\begin{equation}\label{are0}
   \mathcal{P}_0\mathcal{S}_0^T+\mathcal{S}_0\mathcal{P}_0
   -\mathcal{P}_0\mathcal{C}_0^T\mathcal{C}_0\mathcal{P}_0+ I_{pn}=0.
\end{equation}
Let $\alpha_0=\col(\alpha_{0,1},\dots,\alpha_{0,n})$. For $i=1,\dots,N$, let
\begin{equation}\label{nnnado.1}
    \dot{\alpha}_i=\mu_{\alpha}\sum_{j=0}^Na_{ij}(\alpha_j-\alpha_i)
\end{equation}
where $\alpha_i\in \mathbb{R}^n$, $\mu_{\alpha}>0$, and define
\begin{equation} \label{eq12}
    \mathcal{S}_i=\left(
                  \begin{array}{cccc}
                     0&  &  &  \\
                     \vdots&  & I_{n-1} &  \\
                     0&  &  & \\
                    -\alpha_{i,n} & \cdots & -\alpha_{i,2} & -\alpha_{i,1} \\
                  \end{array}
                \right)
    \otimes I_p.
\end{equation}
Note that $(\mathcal{C}_0,\mathcal{S}_i(t))$ is in the observable canonical form and thus is
observable for all $t\geq 0$. Therefore, the following algebraic Riccati equation
\begin{equation}\label{arei}
   \mathcal{P}_i\mathcal{S}_i^T+ \mathcal{S}_i\mathcal{P}_i-\mathcal{P}_i\mathcal{C}_0^T\mathcal{C}_0\mathcal{P}_i+I_{pn}=0
\end{equation}
admits a unique solution $\mathcal{P}_i(t)>0$ for all $t\geq0$.
For $i=1,\dots,N$, let $\mathcal{F}_i=\mathcal{P}_i\mathcal{C}_0^T$ and define the following dynamic compensator
\begin{equation}\label{nnnado.2}
    \dot{\zeta}_i=\mathcal{S}_i\zeta_i+\mu_{\zeta}\mathcal{F}_i\sum_{j=0}^Na_{ij}(y_j-y_i)
\end{equation}
where $\zeta_i\in \mathbb{R}^{pn}$, $y_i=\mathcal{C}_0\zeta_i$, $\mu_{\zeta}>0$.

For $i=1,\dots,N$, let $\tilde{\alpha}_i=\alpha_i-\alpha_0$,
$\tilde{\mathcal{S}}_i=\mathcal{S}_i-\mathcal{S}_0$, $\tilde{\mathcal{P}}_i=\mathcal{P}_i-\mathcal{P}_0$,
$\tilde{\zeta}_i=\zeta_i-\zeta_0$ and $\tilde{y}_i=y_i-y_0$.
We have the following result.

\begin{Theorem}\label{lem3}
Given systems \eqref{leader} and \eqref{nnnado.1}, \eqref{nnnado.2},
under Assumption \ref{ass1},
if $\mu_{\alpha}> \bar{\delta}_{S_0}\underline{\delta}_H^{-1}$ and $\mu_{\zeta}>\underline{\delta}_H^{-1}$,
then for any $v_0(0)\in \mathbb{R}^q$,
$\alpha_i(0)\in \mathbb{R}^n$,
$\zeta_i(0)\in \mathbb{R}^{pn}$, $i=1,\dots,N$,
$\alpha_i(t)$ and $\zeta_i(t)$ exist for
all $t\geq 0$ and satisfy
$\lim_{t\rightarrow\infty}\tilde{\alpha}_i(t)=0$,
$\lim_{t\rightarrow0}\tilde{\mathcal{S}}_i(t)=0$,
$\lim_{t\rightarrow0}\tilde{\mathcal{P}}_i(t)=0$,
$\lim_{t\rightarrow\infty}\tilde{\zeta}_i(t)=0$,
$\lim_{t\rightarrow\infty}\tilde{y}_i(t)=0$.
\end{Theorem}

\begin{Proof}
 Let $\tilde{\alpha}=\col(\tilde{\alpha}_1,\dots,\tilde{\alpha}_N)$. Note that $\alpha_i$ is governed by (\ref{nnnado.1}), which is in the same form as (\ref{ado.1}). Thus, by Part (i) of Lemma \ref{lem1},  under Assumption \ref{ass1},  for any
  $\mu_{\alpha}> 0$, $\tilde{\alpha}_i(t)$ decays to zero exponentially at
  the rate of $\mu_{\alpha}\underline{\delta}_H$,  which together with (\ref{eq12}) implies
 $\tilde{\mathcal{S}}_i(t)$ decays to zero exponentially at
  the rate of $\mu_{\alpha}\underline{\delta}_H$. Note that $\mathcal{P}_i(t)$ is unique
  for all $t\geq 0$ and thus is continuous in $\mathcal{S}_i$ by Lemma \ref{lemap1}. Therefore, $\lim_{t\rightarrow\infty}\tilde{\mathcal{S}}_i(t)=0$ implies that
  $\lim_{t\rightarrow\infty}\tilde{\mathcal{P}}_i(t)=0$.
Moreover, by \eqref{exoeq} and \eqref{nnnado.2}, we have
\begin{equation}
\begin{aligned}
    \dot{\tilde{\zeta}}_i&=\mathcal{S}_i\zeta_i+\mu_{\zeta}\mathcal{F}_i\sum_{j=0}^Na_{ij}(y_j-y_i)
    -\mathcal{S}_0\zeta_0\\
    &=\mathcal{S}_i\zeta_i-\mathcal{S}_0\zeta_i+\mathcal{S}_0\zeta_i-\mathcal{S}_0\zeta_0\\
    &+\mu_{\zeta}\mathcal{P}_i\mathcal{C}_0^T\mathcal{C}_0\sum_{j=0}^Na_{ij}(\zeta_j-\zeta_i)\\
    &=\mathcal{S}_0\tilde{\zeta}_i+\tilde{\mathcal{S}}_i\zeta_i
    +\mu_{\zeta}\mathcal{P}_i\mathcal{C}_0^T\mathcal{C}_0\sum_{j=0}^Na_{ij}(\tilde{\zeta}_j-\tilde{\zeta}_i)\\
    &=\mathcal{S}_0\tilde{\zeta}_i+\tilde{\mathcal{S}}_i\tilde{\zeta}_i+\tilde{\mathcal{S}}_i\zeta_0
    +\mu_{\zeta}\mathcal{P}_0\mathcal{C}_0^T\mathcal{C}_0\sum_{j=0}^Na_{ij}(\tilde{\zeta}_j-\tilde{\zeta}_i)\\
    &+\mu_{\zeta}\tilde{\mathcal{P}}_i\mathcal{C}_0^T\mathcal{C}_0\sum_{j=0}^Na_{ij}(\tilde{\zeta}_j-\tilde{\zeta}_i).
\end{aligned}
\end{equation}
Let $\tilde{\zeta}=\col(\tilde{\zeta}_1,\dots,\tilde{\zeta}_N)$, $\tilde{\mathcal{S}}_d=\hbox{block diag}(\tilde{\mathcal{S}}_1,\dots,\tilde{\mathcal{S}}_N)$
and $\tilde{\mathcal{P}}_d=\hbox{block diag}(\tilde{\mathcal{P}}_1,\dots,\tilde{\mathcal{P}}_N)$. Then
\begin{equation}\label{lem0.1}
\begin{aligned}
   \dot{\tilde{\zeta}}&=\left(I_N\otimes \mathcal{S}_0-\mu_{\zeta} (H\otimes \mathcal{P}_0\mathcal{C}_0^T\mathcal{C}_0)\right)\tilde{\zeta}\\
   &+\left(\tilde{\mathcal{S}}_d-\mu_{\zeta}\tilde{\mathcal{P}}_d(H\otimes \mathcal{C}_0^T\mathcal{C}_0)\right)\tilde{\zeta}+\tilde{\mathcal{S}}_d(1_N\otimes \zeta_{0})\\
   &\triangleq \mathcal{S}_{\alpha}\tilde{\zeta}+\mathcal{S}_{\beta}(t)\tilde{\zeta}+\mathcal{S}_{\gamma}(t)
\end{aligned}
\end{equation}
where $\mathcal{S}_{\alpha}=I_N\otimes \mathcal{S}_0-\mu_{\zeta} (H\otimes \mathcal{P}_0\mathcal{C}_0^T\mathcal{C}_0)$,
$\mathcal{S}_{\beta}(t)=\tilde{\mathcal{S}}_d-\mu_{\zeta}\tilde{\mathcal{P}}_d(H\otimes \mathcal{C}_0^T\mathcal{C}_0)$,
and $\mathcal{S}_{\gamma}(t)=\tilde{\mathcal{S}}_d(1_N\otimes \zeta_{0})$.
Under Assumption \ref{ass1}, $\underline{\delta}_H>0$.
Then, by Lemma \ref{cor2}, $\mathcal{S}_{\alpha}$
is Hurwitz given $\mu_{\zeta}>\underline{\delta}_H^{-1}$. Thus,
system \eqref{lem0.1} is input-to-state stable
  viewing $\mathcal{S}_{\beta}(t)\tilde{\zeta}+\mathcal{S}_{\gamma}(t)$ as the input \cite{Sontag95}. Therefore, it has the asymptotic gain property \cite{Sontag96}, that is, there exists a class $\mathcal{K}$ function
  $\phi$ such that, for any initial condition, $\tilde{\zeta}(t)$ satisfies
  \begin{equation}\label{rag}
    \limsup_{t \rightarrow\infty} ||\tilde{\zeta}(t)|| \leq \phi\left(\limsup_{t \rightarrow\infty}||\mathcal{S}_{\beta}(t)\tilde{\zeta}(t)+\mathcal{S}_{\gamma}(t)|| \right).
  \end{equation}
We now further show  $\lim_{t\rightarrow\infty} \tilde{\zeta}(t)=0$. For this purpose, consider the following system
  \begin{equation}\label{lem0.2}
\dot{\tilde{\zeta}}=\mathcal{S}_{\alpha}\tilde{\zeta}+\mathcal{S}_{\beta}(t)\tilde{\zeta}.
\end{equation}
Since $\mathcal{S}_{\alpha}$ is Hurwitz and $\mathcal{S}_{\beta}(t)\rightarrow0$ as $t\rightarrow\infty$,
by Lemma 1 of \cite{clgh16}, the origin of \eqref{lem0.2}
  is exponentially stable. As a result,
  system \eqref{lem0.1} is input-to-state stable
  viewing $\mathcal{S}_{\gamma}(t)$ as the input \cite{Sontag95}, which implies that
   the solution of \eqref{lem0.1} is bounded for any initial condition.
 Moreover,  since $\mu_{\alpha}> \bar{\delta}_{S_0}\underline{\delta}_H^{-1}$, $\mathcal{S}_{\gamma}(t)$ decays to zero exponentially.
Thus, it follows from  (\ref{rag}) that $\lim_{t\rightarrow\infty} \tilde{\zeta}(t)=0$.
Finally, noting that $\tilde{y}_i=\mathcal{C}_0\tilde{\zeta}_i$ gives $\lim_{t\rightarrow\infty} \tilde{y}_i(t)=0$.

\end{Proof}

\begin{Remark}
  If none of the eigenvalues of $S_0$ have positive real parts,
 then Lemma \ref{lem3} holds
  for any $\mu_{\alpha}>0$.
\end{Remark}

\begin{Remark}
 To achieve the aforementioned two improvements, we have firstly parameterized the system matrix $S_0$ of the leader system by the coefficients of
its minimal polynomial, which, on one hand, reduces the required information of the leader system,
and on the other hand, guarantees that the pair $(\mathcal{C}_0,\mathcal{S}_i(t))$ is always
observable, which in turn guarantees that the solution to the quadratic nonlinear Riccati equation \eqref{arei}
is unique. Then, according to Lemma 2.2, the solution $\mathcal{P}_i(t)$ to the Riccati equation
is continuous in $\mathcal{S}_i(t)$. Thus,
$\lim_{t\rightarrow\infty}\tilde{\mathcal{S}}_i(t)=0$ implies
$\lim_{t\rightarrow\infty}\tilde{\mathcal{P}}_i(t) = 0$, which eventually enables
the design of the certainty equivalent gain $\mathcal{F}_i=\mathcal{P}_i\mathcal{C}_0^T$ of \eqref{nnnado.2}.
\end{Remark}

%\begin{Remark}
%In order to estimate $y_0$ using \eqref{ado}-\eqref{ado.2}, the dimension of the observer is $(q+1+p)\times q$. While,
%in order to estimated $y_0$ using \eqref{nnnado.1}-\eqref{nnnado.2}, the dimension of the observer is
%$(p+1)\times n$. Noting that $n\leq q$, the dimension of the observer using \eqref{nnnado.1}-\eqref{nnnado.2}
%is at least $q^2$ lower than that using \eqref{ado}-\eqref{ado.2}. Moreover, the information exchange between two agents
%using  \eqref{ado}-\eqref{ado.2} is in the dimension of $(q+1+p)\times q$. While, the information exchange between two agents
%using \eqref{nnnado.1}-\eqref{nnnado.2} is in the dimension of $(n+p)$, which is much lower
%than that using \eqref{ado}-\eqref{ado.2}.
%\end{Remark}

\section{Example}\label{ss}

In this section, we illustrate our approach by a numerical example.
Consider a multiagent system of one leader and four followers. The leader system is given by
\begin{subequations}
    \begin{align}
    \dot{v}_0 &=\left(
                  \begin{array}{ccccc}
                    0 & 0 & 0 & 0 & 0 \\
                    0 & 0 & 1 & 0 & 0 \\
                    0 & -1 & 0 & 0 & 0 \\
                    0 & 0 & 0 & 0 & 2 \\
                    0 & 0 & 0 & -2 & 0 \\
                  \end{array}
                \right)v_0=S_0v_0\\
     v_0(0)&=\left(
               \begin{array}{ccccc}
                 1 & 0 & 1 & 0 & 1 \\
               \end{array}
             \right)^T\\
   y_0 &= \left(
                      \begin{array}{ccccc}
                        0.5 & 0 & 0 & 0 & 0 \\
                        0 & 1 & 0 & 0 & 0\\
                        0 & 0 & 0 & 0 & 2\\
                      \end{array}
                    \right)v_0=C_0v_0.
      \end{align}
\end{subequations}

The minimal polynomial of $S_0$ is given by $s^5+5s^3+4s$. Therefore, we have
\begin{subequations}
    \begin{align}
   \mathcal{S}_0 &=\left(
                  \begin{array}{ccccc}
                    0 & 1 & 0 & 0 & 0 \\
                    0 & 0 & 1 & 0 & 0 \\
                    0 & 0 & 0 & 1 & 0 \\
                    0 & 0 & 0 & 0 & 1 \\
                    0 & -4 & 0 & -5 & 0 \\
                  \end{array}
                \right)\otimes I_3\\
   \mathcal{C}_0 &= \left(
                      \begin{array}{ccccc}
                        1 & 0 & 0 & 0& 0 \\
                      \end{array}
                    \right)\otimes I_3\\
   \alpha_0&=\left(
               \begin{array}{ccccc}
                 0 & -5 & 0 & -4 & 0 \\
               \end{array}
             \right)^T.
      \end{align}
\end{subequations}

\begin{figure}
\begin{center}
\scalebox{1}{\includegraphics[125,574][268,728]{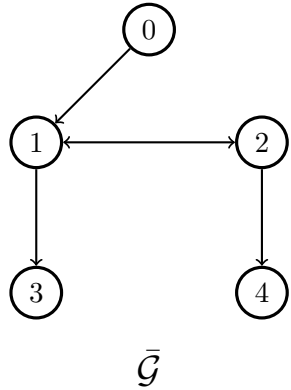}}
\caption{Communication network $\bar{\mathcal{G}}$.}\label{topo}
\end{center}
\end{figure}

\begin{figure}
\begin{center}
\scalebox{0.56}{\includegraphics[0,0][400,300]{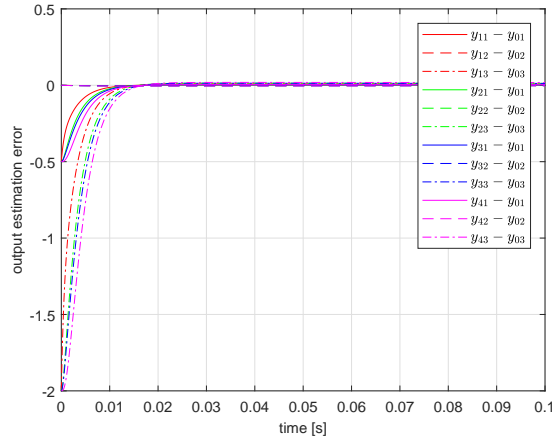}}
\caption{Output estimation performance.}\label{a-1}
\end{center}
\end{figure}

The communication graph $\bar{\mathcal{G}}$ is
shown in Fig. \ref{topo}. The gains of the output based adaptive distributed output observer are given
as $\mu_{\alpha}=10$, $\mu_{\zeta}=200$. We let $a_{ij}=1$ whenever $(j,i) \in \bar{\mathcal{E}}$.
The elements of $\alpha_i(0)$ and $\zeta_i(0)$ for $i=1,2,3,4$ are taken from $[-1,1]$.
Fig. \ref{a-1} shows the output estimation errors of the output based adaptive distributed observers.

It is interesting to make a comparison between the dimension
of the output based adaptive distributed output observer \eqref{nnnado.1}, \eqref{nnnado.2}
and the dimension of
the state based adaptive distributed output observer \eqref{ado}. In fact, simple calculation shows that
the dimension of the state based adaptive distributed output observer \eqref{ado} is
 45 (25 for estimating $S_0\in \mathbb{R}^{5\times 5}$, 15 for estimating
$C_0\in \mathbb{R}^{3\times 5}$, and 5 for estimating $v_0\in \mathbb{R}^5$) while
the dimension of the output based adaptive distributed output observer \eqref{nnnado.1}, \eqref{nnnado.2} is 20
(5 for estimating $\alpha_0\in \mathbb{R}^5$ and
15 for estimating $\zeta_0\in \mathbb{R}^{15}$).
The comparisons between the dimensions as well as the information exchanges of the two observers
are summarized in Table \ref{table1}.

\begin{table}
\centering
\caption{Comparison between state based adaptive distributed output observer
 and output based adaptive distributed output observer (SB: state based; OB: output based).}
\begin{small}
\begin{tabular}{|c|c|c|}
  \hline
   & control law dimension & information exchange \\ \hline
  SB & \begin{tabular}{cc}

                  $S_i\in \mathbb{R}^{5\times 5}$ & 25 \\
                  $C_i\in \mathbb{R}^{3\times 5}$ & 15 \\
                  $v_i \in \mathbb{R}^5$ & 5 \\ \hline
                  total & \textbf{45} \\

                \end{tabular}
   & \begin{tabular}{cc}

                  $S_i\in \mathbb{R}^{5\times 5}$ & 25 \\
                  $C_i\in \mathbb{R}^{3\times 5}$ & 15 \\
                  $v_i \in \mathbb{R}^5$ & 5 \\ \hline
                  total & \textbf{45} \\

                \end{tabular} \\ \hline
  OB & \begin{tabular}{cc}

                  $\alpha_i\in \mathbb{R}^5$ & 5 \\
                  $\zeta_i\in \mathbb{R}^{15}$ & 15 \\ \hline
                  total & \textbf{20} \\

                \end{tabular} & \begin{tabular}{cc}

                  $\alpha_i\in \mathbb{R}^5$ & 5 \\
                  $y_i\in \mathbb{R}^3$ & 3 \\ \hline
                  total & \textbf{8} \\

                \end{tabular} \\
  \hline
\end{tabular}\label{table1}
\end{small}
\end{table}

\section{Conclusion}\label{sc}
In this communique, we have proposed
an output based adaptive distributed output observer.
In contrast to the existing state based adaptive distributed output observers,
the proposed output based adaptive distributed output observer only needs to know the output of the leader
system. In addition, the dimension as well as the information exchange among agents of the output
based adaptive distributed output observer can be significantly reduced in comparison with the
state based adaptive distributed output observer.


\begin{thebibliography}{1}

%\bibitem{at09}
%A. Abdessameud and A. Tayebi, ``Attitude synchronization of a group of spacecraft without velocity
%measurements'', {\it IEEE Transactions on Automatic Control}, vol. 54, no. 11, pp. 2642-2648, 2009.

%\bibitem{baw11}
%H. Bai, M. Arcak and J. Wen, {\it Cooperative control design: a systematic, passivity-based approach},
%Springer, NY, 2011.
%

%\bibitem{ch14}
%H. Cai and J. Huang, ``The leader-following attitude control of multiple rigid
%spacecraft systems'', {\it Automatica}, vol. 50, no. 4, pp.  1109-1115, 2014.

%\bibitem{chijgs14}
%H. Cai and J. Huang, ``Leader-following consensus of multiple uncertain Euler-Lagrange systems
%under switching network topology,'' {\it International Journal of General Systems},
%vol. 43. no. 3-4, pp. 294-304, 2014.

%\bibitem{ch16}
%H. Cai and J. Huang, ``Leader-following adaptive consensus of multiple uncertain rigid spacecraft
%systems'', {\it Science China Information Sciences}, vol. 59, no. 1, pp.  1-13, 2016.

%\bibitem{ch162}
%H. Cai and J. Huang, ``Leader-following attitude consensus of multiple rigid body systems
%by attitude feedback control'', {\it Automatica}, vol. 69, pp. 87-92, 2016.

%\bibitem{ch163}
%H. Cai and J. Huang, ``The leader-following consensus for multiple uncertain Euler-Lagrange
%systems with an adaptive distributed observer'',
%{\it IEEE Transactions on Automatic Control}, vol. 61, no. 10, pp. 3152-3157, 2016.

%\bibitem{chifac17}
%H. Cai and J. Huang,``Leader-following attitude consensus of multiple rigid body systems by an adaptive distributed observer approach,''
%{\it The 20th IFAC World Congress}, Toulouse, France, 9-14,  July 2017.


\bibitem{chwcica18}
H. Cai and J. Huang,``The attitude consensus problem via an output
based adaptive distributed observer approach,''
{\it Proceedings of the 2018 13th World
Congress on Intelligent Control and Automation}, July 4-8, 2018, Changsha, China, 2018.

\bibitem{clgh16}
H. Cai, F. L. Lewis, G. Hu and J. Huang, ``The adaptive distributed observer approach to the cooperative output regulation of linear multiagent systems'', {\it Automatica}, vol. 75, pp. 299-305, 2017.


\bibitem{chen14}
Z. Chen, ``Pattern synchronization of nonlinear heterogeneous
multiagent networks with jointly
connected topologies'', {\it IEEE Transactions on Control of Network Systems}, vol. 1, no. 4,  pp. 349-359, 2014.


\bibitem{fhrd18}
Z. Feng, G. Hu, W. Ren, W. E. Dixon and J. Mei,
``Distributed coordination of multiple unknown Euler-Lagrange Systems'',
\textit{IEEE Transactions on Control of Network Systems},
vol. 5, no. 1, pp. 55-66, 2018.


%\bibitem{bch15}
%Z. Chen and J. Huang, {\it Stabilization and Regulation of Nonlinear Systems
%: A Robust and Adaptive Approach}, Springer, 2015.




%\bibitem{dc16}
%H. Du, M. Z. Q. Chen and G. Wen,
%``Leader-following attitude consensus for spacecraft formation with rigid and flexible spacecraft'',
%{\it Journal of Guidance, Control and Dynamics}, vol. 39, no. 4, pp. 941-948, 2016.
%
%\bibitem{dl11}
%H. Du, S. Li and C. Qian,
%``Finite-Time attitude tracking control of spacecraft with application to attitude synchronization'',
%{\it IEEE Transactions on Automatic Control}, vol. 56, no. 11, pp. 2711-2717, 2011.







%\bibitem{slotine2009}
%S. -J. Chung and J. -J. E. Slotine, ``Cooperative robot control and
%concurrent synchronization of Lagrangian systems,'' {\it IEEE Trans.
%Robotics}, vol. 25, no. 3, pp. 686-700, Jun. 2009.
%
%\bibitem{dix03} W. E. Dixon, A. Behal, D. M. Dawson and S. P. Nagarkatti,
%\textit{Nonlinear Control of Engineering Systems: A Lyapunov-Based Approach},
%Boston: Birkhauser, 2003.

%\bibitem{bk1}
%C. Godsil and G. Royal, \textit{Algebraic Graph Theory}, New York: Springer-Verlag, 2001.

%\bibitem{gza15}
%S. Goh, S. Zekavat and O. Abdelkhalik,
%``LEO satellite formation for SSP: energy and doppler analysis'',
%{\it IEEE Transactions on Aerospace and Electronic Systems}, vol. 51, no. 1, pp. 18-30, 2015.




%\bibitem{hgc2007}
%Y. Hong, L. Gao, D. Cheng and J. Hu, ``Lyapunov-based approach to multiagent systems
%with switching jointly connected interconnection,'' \textit{IEEE Trans.
%Automat. Control}, vol. 52, no. 5, pp. 943-948, May. 2007.

\bibitem{horn}
R. Horn and C. Johnson, {\it Topics in Matrix Analysis}, Cambridge University Press, 1991.

%\bibitem{h16}
%J. Huang, ``Certainty equivalence, separation principle, and cooperative output regulation of
%multiagent systems by the distributed observer approach'', In K. G. Vamvoudakis and S. Jagannathan
%(Eds.), Control of Complex Systems: Theory and Applications. Elsevier, pp. 421-449, 2016.
%
%\bibitem{h17}
%J. Huang, ``Adaptive distributed observer and the cooperative control of multiagent systems'',
%{\it Journal of Control and Decision}, vol. 4, no. 1, pp. 1-11, 2017.


%{\color{blue}\bibitem{hhg06}
%Y. Hong, J. Hu and L. Gao, ``Tracking control for multi-agent consensus with an active
%leader and variable topoloty'', \textit{Automatica}, vol. 42, no. 7, pp. 1177-1182, 2006.
%
%\bibitem{hcb08}
%Y. Hong, G. Chen and L. Bushnell, ``Distributed observers design for leader-following control of
%multi-agent networks'', \textit{Automatica}, vol. 44, no. 3, pp. 846-850, 2008.}



\bibitem{Hu1}
J. Hu and Y. Hong, ``Leader-following coordination of multiagent
systems with coupling time delays,'' \emph{Physica A: Statistical Mechanics
and its Applications}, vol. 374, no. 2, pp. 853--863, 2007.

\bibitem{hdlr20}
Y. Hua, X. Dong, Q. Li and Z. Ren,
``Distributed adaptive formation tracking for heterogeneous multiagent systems
with multiple nonidentical leaders and without well-informed follower'',
\textit{International Journal of Robust and Nonlinear Control}, vol. 30, pp. 2131-2151, 2020.


\bibitem{jlm03}
A. Jadbabaie, J. Lin, and A. S. Morse, ``Coordination of groups of
mobile agents using nearest neighbor rules,'' \textit{IEEE Transactions on
Automatic Control}, vol 48, no. 6, pp. 988-1001, Jun. 2003.

%\bibitem{kha02}
%H. K. Khalil, {\it Nonlinear Systems}, Prentice Hall, Upper Saddle River, NJ, 3rd edition, 2002.
%

\bibitem{kc16}
S. Knorn, Z. Chen, R. Middleton, ``Overview: collective control of multiagent systems,'' \textit{IEEE Transactions on
Control of Network Systems}, vol 3, no. 4, pp. 334-347, Jun. 2016.

%\bibitem{k72}
%V. Ku\v{c}era, ``A contribution to matrix quadratic equations'',
%{\it IEEE Transactions on Automatic Control}, vol. 17, no. 3, pp. 344-347, 1972.

%\bibitem{lewis} F. L. Lewis, S. Jagannathan and A. Yesildirek,
%\textit{Neural Network Control of Robot Manipulatros and Nonlinear Systems}, Taylor and
%Francis, 1999.
\bibitem{lr95}
P. Lancaster, L. Rodman, {\it Algebraic Riccati equations}, Oxford University Press, 1995.

\bibitem{lzd14}
F. L. Lewis, H. Zhang, K. Hengster-Movric, A. Das,
{\it Cooperative Control of Multi-Agent Systems: Optimal and Adaptive Design Approacheds},
Springer-Verlag: London, UK, 2014.

\bibitem{lzmz19}
H. Liang, Y. Zhou, H. Ma and Q. Zhou, ``Adaptive distributed observer approach for
cooperative containment control of nonidentical networks'',
\textit{IEEE Transactions on Systems, Man, and Cybernetics: Systems}, vol. 49, no. 2, pp. 299-307, 2019.

%\bibitem{lcdl18}
%K. Liu, Y. Chen, Z. Duan, J. Lu, ``Cooperative output regulation of LTI plant via distributed
%observers with local measurement'', \emph{IEEE Transactions on Cybernetics}, vol. 48, no. 7, pp. 2181-2191, 2018.

%\bibitem{lc14}
% H. Liu, C. D. Persis and M. Cao, ``Robust decentralized output regulation with single or multiple
%reference signals for uncertain heterogeneous systems'', {\it International Journal of Robust and Nonlinear
%Control}, vol. 25, no. 9, pp. 1399-1422, 2015.


\bibitem{om04}
R. Olfati-Saber, R. M. Murray, ``Consensus problems in networks of agents with switching
topology and time-delays'', {\it IEEE Transactions on Automatic Control}, vol. 49, no. 9, pp. 1520-1533, 2004.

\bibitem{q09}
Z. Qu, \textit{Cooperative Control of Dynamical Systems: Applications to Autonomous Vehicles}.
Springer-Verlag: London, U.K., 2009.



%\bibitem{lds14}
%S. Li, H. Du and P. Shi, ``Distributed attitude control for multiple spacecraft with communication delays'',
%{\it IEEE Transactions on Aerospace and Electronic Systems}, vol. 50, no. 3, pp. 1765-1773, 2014.

%\bibitem{mmc15}
%G. Mushet, G. Mingotti, C. Colombo and C. McInnes,
%``Self-organising satellite constellation in geostationary earth orbit'',
%{\it IEEE Transactions on Aerospace and Electronic Systems}, vol. 51, no. 2, pp. 910-923, 2015.


%\bibitem{pkg15} C. Pang, A. Kumar, C. Goh and C. Le, ``Nano-satellite swarm
%for SAR applications: design and robust scheduling'',
%{\it IEEE Transactions on Aerospace and Electronic Systems}, vol. 51, no. 2, pp. 853-865, 2015.

%\bibitem{r07}
%W. Ren, ``Distributed attitude alignment in spacecraft formation flying'',
%{\it International Journal of Adaptive Control and Signal Processing}, vol. 21, no. 2-3, pp. 95-113, 2007.

%\bibitem{r072}
%W. Ren, ``Formation keeping and attitude alignment for multiple spacecraft through local
%interactions'', {\it Journal of Guidance, Control and Dynamics}, vol. 30, no. 2, pp. 633-638, 2007.
%

\bibitem{rb08}
W. Ren, R. W. Beard,
\textit{Distributed Consensus in Multi-Vehicle Cooperative Control, Communications and Control
Engineering Series}. Springer-Verlag: London, U.K., 2008.


%\bibitem{sk11}
%R. Schlanbusch, R. Kristiansen and P. J. Nicklasson,
%``Spacecraft formation reconfiguration with collision avoidance'',
% {\it Automatica}, vol. 47, no. 7, pp. 1443-1449, 2011.

%\bibitem{ts14}
%J. Thunberg, W. Song, E. Montijano, Y. Hong, X. Hu,
%``Distributed attitude synchronization control of multiagent systems with switching topologies'',
%{\it Automatica}, vol. 50, no. 3, pp. 832-840, 2014.



\bibitem{Sontag95}
E. D. Sontag, ``On the input-to-state stability property'',
{\it European Journal of Control}, vol. 1, pp. 24-36, 1995.



\bibitem{Sontag96}
E. D. Sontag and Y. Wang, ``New characterizations of input-to-state stability property'',
{\it IEEE Transactions on Automatic Control}, vol. 41, pp. 1283-1294, 1996.

%\bibitem{tuna}
%S. E. Tuna, ``LQR-based coupling gain for synchronization of
%linear systems'', 	arXiv:0801.3390 [math.OC].


%\bibitem{bk2}
%J. -J. E. Slotine and W. Li, \textit{Applied Nonlinear Control}, Prentice-Hall, Englewood Cliffs, NJ, 1991.

\bibitem{ys12}
Y. Su and J. Huang, ``Cooperative output regulation of linear multiagent systems'',
{\it IEEE Transactions on Automatic Control}, vol. 57, no. 4, pp. 1062-1066, 2012.

%\bibitem{ys1}
%Y.  Su and J. Huang, ``Stability of a class of linear switching systems
%with applications to two consensus problems,'' \textit{IEEE Trans.
%Automat. Control}, vol 57, no. 6, pp. 1420-1430, Jun. 2012.
%

\bibitem{ys13} Y. Su, Y.  Hong,  and J. Huang, ``A general result on the cooperative robust output regulation for linear uncertain multiagent systems,''
\textit{IEEE Transactions on
Automatic Control}, vol. 58, no. 5,  pp. 1275-1279, May 2013.


%\bibitem{ys122} Y. Su and J. Huang,
%``Cooperative output regulation with application to multiagent
%consensus under switching network,'' {\it IEEE Trans. on Systems,
%Man, and Cybernetics, Part B: Cybernetics}, vol. 42, no. 3, pp. 864-875, Jun. 2012.

%\bibitem{ys15}
%Y. Su and J. Huang,
%``Cooperative global robust output regulation
%for nonlinear uncertain multiagent systems
%in lower triangular form'',
%{\it IEEE Transactions on Automatic Control}, vol. 60, no. 9, pp. 2378-2389, 2015.


%\bibitem{vh06}
%M. C. VanDyke and C. D. Hall, ``Decentralized coordinated attitude control within a formation of
%spacecraft'', {\it Journal of Guidance, Control and Dynamics}, vol. 29, no. 5, pp. 1101-1109, 2006.
%
%
%\bibitem{wkd91}
%J. T. Wen and K. Kreutz-Delgado, ``The attitude control problem'',
%{\it IEEE Transactions on Automatic Control}, vol. 36, no. 10, pp. 1148-1162, 1991.

%\bibitem{wh96}
%P. K. C. Wang and F. Y. Hadaegh, ``Coordination and control of multiple microspacecraft moving
%in formation'', {\it Journal of Astronautical Sciences}, vol. 43, no. 3, pp. 315-355, 1996.

%\bibitem{y88}
%J. S. -C. Yuan, ``Closed-loop manipulator control using quaternion feedback'',
%{\it IEEE Journal of Robotics and Automation}, vol. 4, no. 4, pp. 434-440, 1988.
%
%
%\bibitem{zcg17}
%C. Zhong, Z. Chen and Y. Guo, ``Attitude control for flexible
%spacecraft with disturbance
%rejection'',
%\textit{IEEE Transactions on Aerospace and
%Electronic Systems}, vol. 53, no. 1, pp. 101-110, 2017.
%
%
%\bibitem{zk12}
%A. Zou and K. Kumar, ``Distributed attitude coordination control for spacecraft formation flying'',
%\textit{IEEE Transactions on Aerospace and
%Electronic Systems}, vol. 48, no. 2, pp. 1329-1346, 2012.

\end{thebibliography}
\end{document}